# Effect of Outlier Removal from Temporal ASF Corrections on Multichain Loran Positioning Accuracy

Jongmin Park[1], Pyo-Woong Son[2], Woohyun Kim[1], Joon Hyo Rhee[3], and Jiwon Seo[1*]

[1] School of Integrated Technology, Yonsei University,
Incheon, 21983, Korea (jm97, crimy00, jiwon.seo@yonsei.ac.kr)
[2] Korea Research Institute of Ships and Ocean Engineering,
Daejeon, 34103, Korea (pwson@kriso.re.kr),
[3] Korea Research Institute of Standards and Science,
Daejeon, 34113, Korea (jh.rhee@kriss.re.kr)
* Corresponding author

**Abstract**: The widely used global navigation satellite systems (GNSSs) are vulnerable to radio frequency interference (RFI). Long-range navigation (Loran), a terrestrial navigation system, can compensate for this weakness; however, it suffers from low positioning accuracy, and studies are under way to improve its positioning performance. One such study has proposed the multichain Loran positioning method that uses the signals of transmitting stations belonging to different chains. Although the multichain Loran positioning performance is superior to the performance of conventional methods, the additional secondary factor (ASF) can still degrade its positioning accuracy. To mitigate the effects of temporal ASF, which is one of the ASF components, it is necessary to obtain temporal correction data from a nearby reference station at a known location. In this study, an experiment is performed to verify the effect of removing the outliers in the temporal correction data on the multichain Loran positioning accuracy.

## 1. INTRODUCTION

As the usage of the global navigation satellite systems (GNSSs) is increasing rapidly, the weakness of GNSSs is also being actively studied [1-7]. Long-range navigation (Loran) is a radio navigation system that utilizes high-power long-wavelength terrestrial signals transmitted from a transmitting station on the ground [8]. Unlike GNSSs, Loran is known to be robust against jamming from a portable jammer [9]. As a result, Loran and enhanced Loran (eLoran) have been proposed to complement the vulnerabilities of GNSSs. However, the current Loran system has very low positioning accuracy.

The multichain Loran positioning [10] is a technique that exhibits improved Loran performance, by using signals of transmitting stations that are not in the same chain. The main error source of the multichain Loran as well as the conventional Loran is signal delay due to land path [11], which is called additional secondary factors (ASF) [12-15]. Thus, ASF correction is essential to improve the Loran positioning performance.

ASF has spatial and temporal components. To correct the spatial ASF, it is usual to prepare ASF maps that measure and record the spatial ASF in advance [16-19]. The ASF map can be produced by calculating the spatial ASF by performing positioning at a point where the actual position is known. To calibrate the temporal ASF, it is necessary to deliver the temporal ASF correction information from a differential correction station to the receiver [20-22]. In this paper, we conducted an experiment to evaluate the effects of outliers in the temporal ASF corrections on the multichain Loran positioning accuracy.

## 2. LORAN MEASUREMENTS AND OUTLIER REMOVAL

### 2.1 Loran measurement data

The experiment was conduction at the Siheung Gaetgol Ecology Park, Korea, whose latitude and longitude were 37.3907 and 126.7789 degree, respectively. Loran signals were collected from 17:15 to 17:45 on May 13, 2019 using the software developed in [23]. The spatial ASF was corrected using the signals collected from the same location, and the temporal ASF was corrected using the data measured by the differential correction station located at Yonsei University, Incheon, Korea, which was approximately 10 km from the receiver location. The latitude and longitude of the correction station were 37.3818 and 126.6702 degree, respectively.

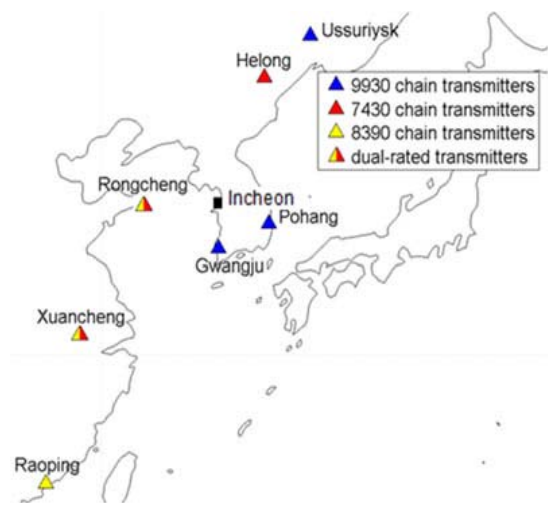

Fig. 1 Loran chains in Northeast Asia (reproduction of Fig. 2 of [10]).

Seven Loran transmitters were used for positioning; including Helong (7430Y), Rongcheng (7430M, 8390Y), Xuancheng (7430X, 8390M), and Raoping (8390X) in China; Pohang (9930M) and Gwangju (9930X) in Korea; and Ussuriysk (9930Z) in Russia, as shown in Fig. 1 [10]. However, Rongcheng and Xuancheng are dual-rated transmitters, belonging to the both 7430 and 8390 chains; therefore, nine types of information were received.

**2.2 Outlier removal**

The "TOA" that is conventionally displayed on Loran receivers is different from the time of arrival (TOA), which is the time taken for the Loran signal to arrive. The "TOA" displayed on the receiver is the remainder of TOA divided by the group repetition interval (GRI), which is the interval between one master transmission and the next. Therefore, "TOA" always has a value between 0 and GRI. To distinguish it from the actual TOA, in this paper, the "TOA" displayed on the receiver is referred to as the time of reception (TOR) following the convention in [10].

Temporal ASF corrections were calculated using the TOR measurements from the correction station. The outliers in the TOR measurements were removed using MATLAB's built-in function "isoutlier." We created temporal ASF corrections using the averages of the temporal ASF over the past five minutes and updated them every minute.

## 3. RESULTS

**3.1 Number of removed outliers**

For each transmitter designator, 432 TOR measurements over 36 min were collected at the differential correction station. The number of outliers of the TOR measurements for each transmitter designator is shown in Table 1; there were 7.22 outliers (i.e., 1.7% of total measurements), on an average in the 36-min data. (Note that a dual-rated transmitter has two transmitter designators.)

Table 1 Number of outliers of TOR for each transmitter designator.

|  | 7430 | | | 8390 | | | 9930 | | |
|---|---|---|---|---|---|---|---|---|---|
|  | M | X | Y | M | X | Y | M | W | Z |
| Num of outliers | 5 | 6 | 8 | 6 | 2 | 10 | 19 | 3 | 6 |

**3.2 Effect of outliers on the positioning performance**

Figs. 2 and 3 and Table 2 show the positioning performances corrected using the temporal ASF calculated without removing the outliers of TOR and the temporal ASF calculated after removing the outliers of TOR. The 95th percentile errors were 12.28 and 12.09 m, with a difference of 0.19 m. However, the difference between the 99th percentile errors was 0.79 m, and the difference in the worst case was 1.88 m. This result is understandable because 1.7% of data were outliers on average, which has little impact on the 95% accuracy but 99% accuracy is affected.

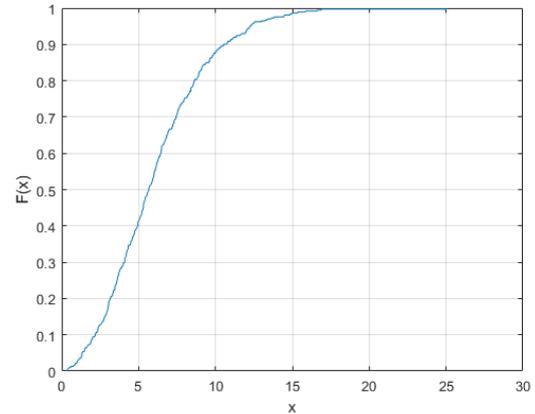

Fig. 2 CDF of positioning result corrected with temporal ASF calculated without removing outliers.

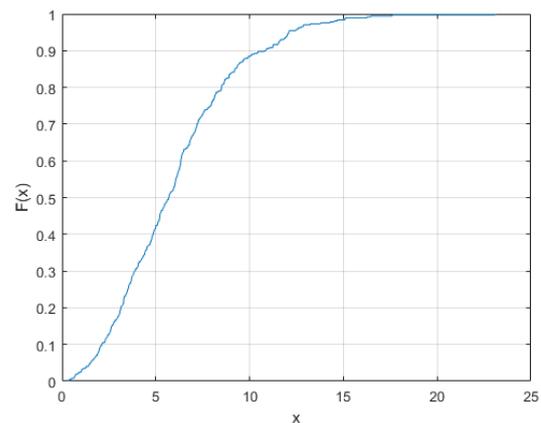

Fig. 3 CDF of positioning result corrected with temporal ASF calculated after removing outliers.

Table 2 Positioning accuracies of cases where outliers were not removed and those where outliers were removed.

|  | 95% accuracy | 99% accuracy | 100% accuracy (i.e., worst case) |
|---|---|---|---|
| Outliers remained | 12.28 m | 15.89 m | 25.03 m |
| Outliers removed | 12.09 m | 15.1 m | 23.11 m |

## 4. CONCLUSIONS

In this paper, we analyzed the effects of the outliers in the TOR measurements on the Loran positioning accuracy. The 36-min TOR data collected at the differential correction station were used to generate the temporal ASF corrections. The 95% accuracy of the receiver was improved slightly by 0.19 m, the 99% accuracy was improved by 0.79 m, and the worst case error was improved by 1.88 m after the outlier removal.

## ACKNOWLEDGEMENT

This research was a part of the project titled "Development of enhanced Loran system," funded by the Ministry of Oceans and Fisheries, Korea.